\documentstyle[sprocl]{article}

\input{psfig}

\def\be{\begin{equation}}
\def\ee{\end{equation}}
\def\bea{\begin{eqnarray}}
\def\eea{\end{eqnarray}}
\begin{document}

\title{THE ANOMALOUS TRAJECTORIES OF THE \\ PIONEER 
SPACECRAFT\footnote{Electronic addresses:  
$^\ddagger$mmn@lanl.gov ~~
$^*$john.d.anderson@jpl.nasa.gov~~ \\
$^+$Philip.A.Laing@aero.org~~  
$^\S$Eunice.L.Lau@jpl.nasa.gov ~~
$^\dagger$turyshev@jpl.nasa.gov
}}

\author{Michael Martin Nieto,$^\ddagger$
John D. Anderson,$^*$ Philip A. Laing,$^+$ \\
Eunice L. Lau,$^\S$ 
and Slava G. Turyshev$^\dagger$\\
~~\\}

\address{
$^\ddagger$Theoretical Division (MS-B285), Los Alamos National Laboratory, \\
University of California, Los Alamos, NM 87545 \\
$^{*,\S,\dagger}$Jet Propulsion Laboratory, California Institute
of  Technology, \\ Pasadena, CA 91109 \\ 
$^+$The Aerospace Corporation, 2350 E. El Segundo Blvd., \\
El Segundo,  CA 90245-4691  
}


\maketitle\abstracts{ 
Because of their unique designs, the Pioneer 10 and 11 spacecraft 
have provided the cleanest Doppler, deep-space navigation data.  Analysis
of this data can be interpreted as showing an anomalous acceleration of
these craft directed towards the Sun of  
 $ a_P \sim  8 \times 10^{-8} ~~{\rm cm/s}^2$. 
The background of this discovery and 
the significance of the result are discussed. 
}

%**********************************

\section{Introduction}

Some thirty years ago, on 2 March  1972, Pioneer 10 was launched on
an Atlas/Centaur rocket from  Cape Canaveral.  Pioneer 10 was Earth's first
space probe to reach an outer planet.  After surviving intense radiation,
 on 4 December 1973  it successfully   encountered 
Jupiter \cite{science}$^{-~}$\cite{pioweb}.  

Today we are all used to spectacular photographs from
the solar system.  But in 1973 it was different.   
For those of us (speaker - MMN) who remember, the
impact of the first Jupiter encounter photographs 
was astounding.  To understand this one only has to 
compare the 25 January 1974 cover of Science~\cite{science} 
with the best Palomar telescope photographs.
 
\begin{figure}[t]  
\psfig{figure=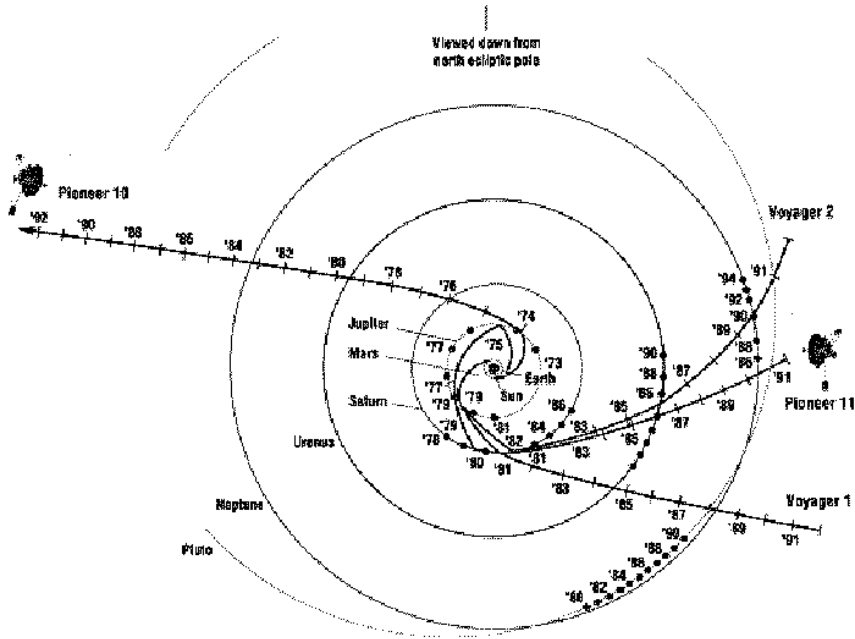,height=90mm}
  \caption{Ecliptic pole view of Pioneer 10, Pioneer 11, 
and Voyager trajectories.  
Pioneer 11 is traveling approximately in the direction of the 
Sun's orbital motion about the galactic center.  The galactic center 
is approximately in the direction of the top of the figure.   
% [Digital artwork by T. Esposito. NASA ARC Image \# AC97-0036-3.]
 \label{fig:pioneerpath}}
\end{figure}

Pioneer 10 was followed by  Pioneer
11 (launched on  5 April 1973).   After Jupiter and 
(for Pioneer 11) Saturn encounters, the two spacecraft 
have followed hyperbolic
orbits near  the plane of the ecliptic to  opposite sides of the solar
system.  Pioneer 10 was also the first mission  to  ``leave the solar
system'' when  in June 1983 it passed beyond the orbit of 
Pluto \cite{extended}.  (See Figure \ref{fig:pioneerpath}.)

%************

%**************

The Pioneers have been amazingly robust.  {\it They knew how to build 
cars in those days!}  The Pioneers were 
adventures into the unknown, so great care 
was given to the design and reliability of the craft.  Hence,  
although  it was required (hoped) these craft would have a lifetime of 
3 (7) years, they successfully operated for 
much longer.  Pioneer 10 (at a distance $> 70$ AU) is still
transmitting and we have analyzed data up to 1998.5. 
The Pioneer 11  Doppler signals failed on 1 October 1990.  So,  after that
date, when the spacecraft was $\sim 30$ AU away from the Sun, no useful
data has been generated for our purposes. 

%********************************************************************

\section{Testing Newtonian Dynamics}

I myself (MMN) became involved in this project while I was preparing a talk 
on antimatter and gravity for the 1994 Low Energy Antiproton conference 
\cite{bled}.  I wanted to argue that  we
really do not understand gravity at distances much larger than the size of
the solar system.  We know it is there, but we don't know how well it obeys
Newton/Einstein physics.  Basically the rotation curves of spiral galaxies
tell us something is wrong.  Our ``creationist'' solution is dark
matter (although we can't find enough of it to explain the problem).  

To better understand what spacecraft could tell us about this, I 
eventually was 
directed to John Anderson.  He emailed me that rather than
considering the orbits of three-axis stabilized craft like the Voyagers,
spinning craft like the Pioneers were better since fewer maneuvers
are necessary.   

How you obtain a result from spacecraft is very simple to understand  --  
in principle.  
You transmit an S-band signal to
the craft via the Deep Space Network. It is transponded back and you
compare the re-received signal to your station clocks. There will be a
Doppler shift since the spacecraft is traveling out.  From this shift 
you subtract out known effects, like the gravity of the Sun and any nearby
planets.\footnote{I am sympathetic to the story Eric Adelberger
told us about the commercial miscalibration of 
their micrometer.  We spent a couple of months
trying to understand  evidence of an anomalously large corona signal.
Eventually we realized that we needed to  input physical 
parameters that had been determined elsewhere instead of using default
operations of the codes.}  Then you see if any Doppler residuals are left
over.

%***********************************************************

\section{The Anomaly}

As to what the Pioneers could actually tell us, 
John  went on, 
``By the way, the biggest systematic in our acceleration residuals
is a bias of $8 \times 10^{-13}$ km/s$^2$ directed toward the Sun.''  
It turns out that indicative evidence of this 
had been around for some time.  But the motivation to analyze it 
seriously had not arisen.  Now it did \cite{anderson}.   

Our collaboration expended a great deal of effort in detailed  
analysis of the Pioneer 10 data between 1987.0 and 1998.5 and of 
the Pioneer 11 data between 1987.0 and 1990.75.  When systematics had
been considered, we obtained a present result  
of \cite{usnew}  $ a_P = (8.74 \pm 1.25) \times 10^{-8} ~~{\rm cm/s}^2$.  

%********************************************************************

\section{A Standard Physic Explanation}

Most people (including ourselves) believe the answer must be some systematic. 
But we have not found it.  The most likely candidates are gas leaks from
the thrusters or non-isotropic heat radiation.  In what probably shows a
deep respect for what they deal with, space navigators tend to think it is
gas leaks and space scientists tend to think it is heat.  

The navigators argue their case privately and the space scientists argue 
their case in print.  The navigational and heat details 
are in our big report \cite{usnew}.   The 
ongoing heat discussions are in our 
references \cite{murphyplus}$^{-~}$\cite{usscheffer}.  
You people will have to decide for 
yourselves, but we claim no ``smoking gun'' has been found.  

%************   

\begin{figure}[t]
\psfig{figure=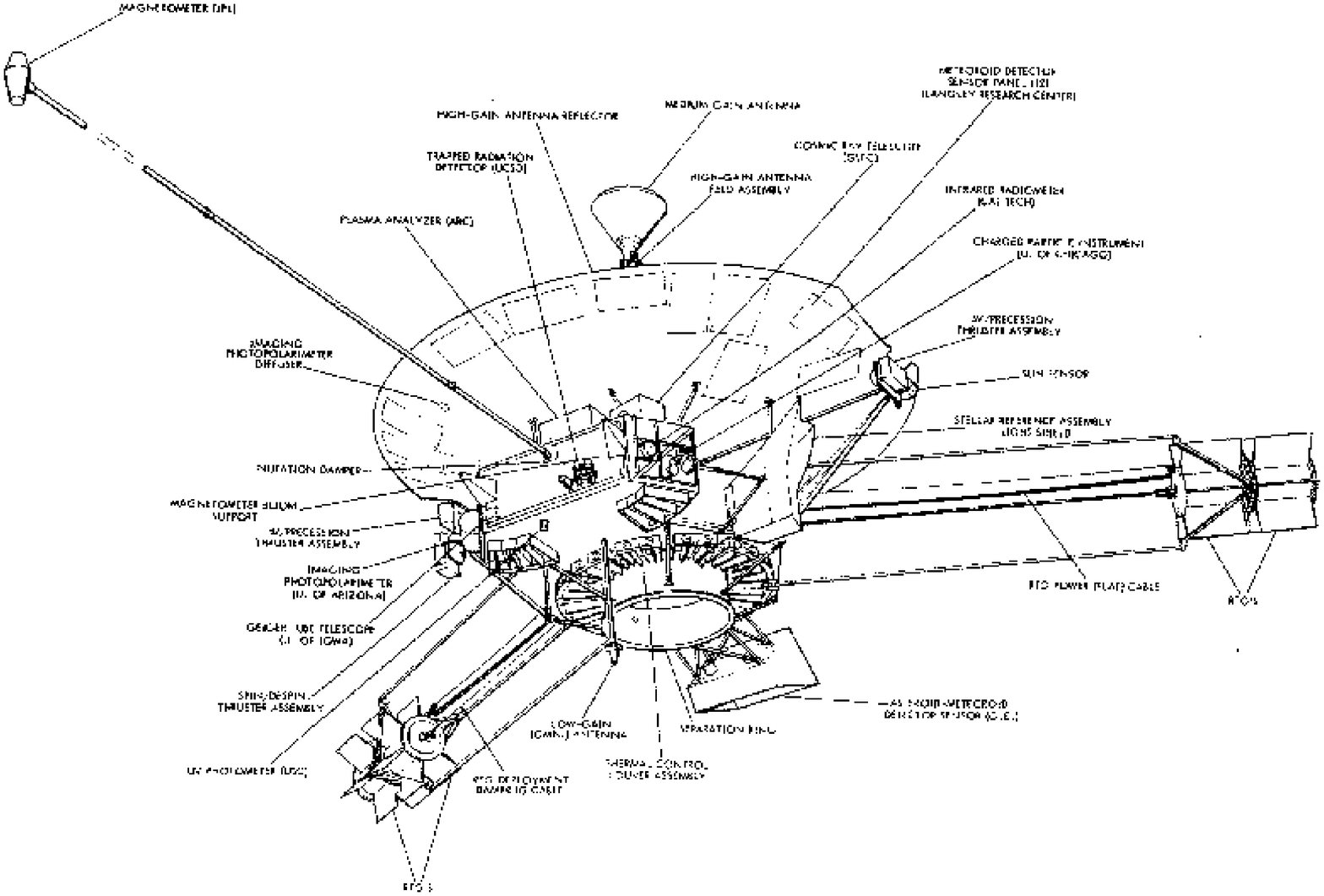,height=3in}
  \caption{A drawing of the Pioneer spacecraft.  
 \label{fig:trusters}}
\end{figure} 

%**************

The care of the design turned out to be beneficial to
us.  (See Figure \ref{fig:trusters}.)
The radioisotope
thermoelectric generators (RTGs) were placed at the end of long booms.  
Therefore, the effect of the radiant heat from them was much less than
it might have been.  Further, 
these extended booms had to be balanced as part of a rotating craft.
This spin-stabilization meant there were fewer maneuvers. 
Finally, the basically  cylindrical symmetry meant one could easily look
for changes in the anomaly with time that might be associated with 
radioactive or electrical-power heat decay. 

%***********************************************

\section{Something Wrong?}

If,  contrary to expectation, no systematic can explain the
anomaly, one has to ask what it could be.  The obvious answer 
to explain an ``acceleration'' is a ``force.''  But as discussed 
elsewhere \cite{anderson,usnew}, 
this force would not satisfy the Principle of Equivalence.  
Still one has to remain curious since 
$a_P \sim {\cal{O}}(a_0) \sim {\cal{O}}(cH)$, where 
$a_0$ is Milgrom's acceleration parameter in his Modified 
Newtonian Dynamics \cite{milg} and $H$ is the Hubble constant.

This last  ties into this conference on CPT.  
One can easily speculate  (and people have)
that the effect is time.
Remember, this is a Doppler measurement which we interpreted as an
acceleration, $a_P$.  But if one writes it as $c a_t = a_P$ one has a time
acceleration. 

The question probably must be settled by new experiment.  
A craft going quickly to deep space, built to minimize systematics, and 
with modern Doppler and range electronics is what is called for. 
We are thinking about this. 

%***************************************************************

\section*{Acknowledgments}

M.M.N. acknowledges support  by the U.S. DOE.
This work of J.D.A., E.L.L, and S.G.T. was performed  under contract
with the  National Aeronautics and Space Administration. 
P.A.L. was supported by a grant from NASA through the
Ultraviolet, Visible, and  Gravitational Astrophysics Program.  

%*************************************************

\end{document}